\documentclass[a4paper,fleqn]{cas-sc}

\usepackage[english]{babel}
\usepackage[round]{natbib}
\usepackage{float}

\def\tsc#1{\csdef{#1}{\textsc{\lowercase{#1}}\xspace}}
\tsc{WGM}
\tsc{QE}
\tsc{EP}
\tsc{PMS}
\tsc{BEC}
\tsc{DE}
\begin{document}
\let\WriteBookmarks\relax
\def\floatpagepagefraction{1}
\def\textpagefraction{.001}

\title [mode = title]{Urban Housing Prices and Migration's Fertility Intentions : Based on the 2018 China Migrants' Dynamic Survey}    

\tnotemark[1]
\tnotetext[1]{The authors thank the Migrant Population Service Center, National Health Commission P. R. China for data support.}

\author[1]{Jingwen Tan}[orcid=0000-0002-3452-959X]
\cormark[2]
\ead{tjw@henu.edu.cn}
\credit{Conceptualization of this study, Methodology, Software}
\affiliation[1]{organization={School of Economics, Henan University},
    addressline={Jinming Avenue}, 
    city={Kaifeng},
    postcode={475004}, 
    country={China}}
\fnmark[1]

% Secend author
\author[1]{Shixi Kang}[orcid=0000-0003-4847-5171]
\cormark[1]
\ead{ksx@henu.edu.cn}
\credit{Conceptualization of this study, Methodology, Software}
% Footnote of the secend author
\fnmark[1]

\cortext[cor1]{Corresponding author}
\cortext[cor2]{Principal Corresponding author}
\fntext[fn1]{These authors contributed equally to this work.}
\begin{abstract}
While the size of China's mobile population continues to expand, the fertility rate is significantly lower than the stable generation replacement level of the population, and the structural imbalance of human resource supply has attracted widespread attention. This paper uses LPM and Probit models to estimate the impact of house prices on the fertility intentions of the mobile population based on data from the 2018 National Mobile Population Dynamics Monitoring Survey. The lagged land sales price is used as an instrumental variable of house price to mitigate the potential endogeneity problem. The results show that for every 100\% increase in the ratio of house price to household income of mobile population, the fertility intention of the female mobile population of working age at the inflow location will decrease by 4.42\%, and the marginal effect of relative house price on labor force fertility intention is EXP(-0.222); the sensitivity of mobile population fertility intention to house price is affected by the moderating effect of infrastructure construction at the inflow location. The willingness to have children in the inflow area is higher for female migrants of working age with lower age, smaller family size and higher education. Based on the above findings, the study attempts to provide a new practical perspective for the mainline institutional change and balanced economic development in China's economic transition phase.
\end{abstract}

% Keywords
\maketitle

\section{Introduction}

In recent years, China's total fertility rate is between 1.5 and 1.6, which is significantly lower than the level of generation replacement for population stabilization and close to the internationally recognized "low fertility trap" threshold of 1.3 (Li, Y. F. et al., 2021). At the same time, the size of China's floating population is expanding, and the rural outflow population provides important human resources for urban development and plays a huge role in the urbanization process (K.H. Zhang, S. Shunfeng, 2003). As of 2020, the number of China's floating population is as high as 375.82 million, an increase of 69.73\% in 10 years (National Bureau of Statistics, 2021). The potential structural imbalance in human resource supply caused by low fertility rate and high labor demand has caused widespread concern.

At the beginning of reform and opening up, the rural migrant population in China was mainly seeking employment; in the 1990s, it changed to mainly seeking increased income; at this time, the migrant population was mostly young adults, showing a trend of "single-handedness", and the number of elderly migrant population and child migrant population were relatively small (Duan, Chengrong et al., 2013). In recent years, the mobility of the population has entered a state where several generations of the population move together, and family-based units have gradually become common (Fan.C.C,Li.T.J,2020). As the structure of the mobile population changes, the dependence of the mobile population on traditional social relationships gradually decreases, and problems such as housing, education, and retirement gradually emerge (Ma, Z.L.Z. 2004). Obviously, because of family reasons, the mobile population has a stronger need to settle than before. However, due to China's traditional dualistic urban-rural household registration system, migrant populations often do not receive local medical and educational resources well, making it difficult for them to integrate socially from an institutional perspective and affecting their fertility intentions (Tu, Y.D, 2020; Hao, P.T, 2015). In the process of urbanization in China, the integration level of the migrant population in terms of social insurance, cultural life, and identity is low, and there is a gap in the allocation of public resources with the residents of the inflowing areas (Chen, Y.S.; Zhang, Y., 2015), resulting in the failure of the migrant population to truly integrate into urban society, and some scholars have proposed the "semi-urbanization Some scholars have proposed the concept of "semi-urbanization" (Wang, 2006).

Alleviating the structural imbalance of human resource supply, promoting the increase of fertility intention of the mobile population, and creating a stable labor source for economic development is one of the important paths to promote China's advancement to a high-income country. The exploration of the fertility intention of mobile population from the perspective of housing price is not only related to the coordinated development of regional economy, the reduction of social wealth gap and the alleviation of social conflicts, but also can optimize the allocation of regional labor endowment structure and promote the stable and healthy development of China's economy.

\section{Research background}

The study of human capital is an important part of neoclassical economics \citep{dettling2014house}. Whether it is aimed at the micro level of increasing individual well-being and improving the family environment \citep{petch2008psycho} or the macro perspective based on promoting the reproduction of human capital and thus the proper functioning of social institutions \citep{gough2017birth}, the study of fertility intentions has obvious relevance. A number of studies have already explored this topic from a multidimensional perspective. For example, \cite{kim2019relationship} found in a study of low fertility in Korea that socio-environmental factors can influence fertility intentions through individual environmental factors. As another example, \cite{arai2007peer} further argues for the indirect role of social relationships on fertility intentions through an empirical analysis of the potential impact of community on fertility intentions. Among them, studies using house prices as a source of influence are more extensive. \cite{dettling2014house} define fertility prices as housing costs and argue that the real estate market has a more direct effect on fertility intentions. \cite{lino2002expenditures} finds a significant negative relationship between house prices and fertility intentions, suggesting that housing costs are the largest expense of raising children.

Regarding the mechanism of the effect of housing prices on the fertility intention of the mobile population, domestic and foreign research focuses on both the willingness to integrate and the financial ability. On the one hand, the high or low housing price is related to the strength of the social integration intention of the mobile population. According to \cite{entzinger2003benchmarking}, social integration includes psychological integration, cultural integration, economic integration, and political integration. House price, as a specific quantification of property rights, is an important indicator of the mobile population's integration into the society of the place of migration in many ways, which in turn affects the willingness to stay. Since the migration of the mobile population is a short-term decision-making behavior\citep{duyang2014}, lowering the threshold of ownership of their property rights is a key initiative to promote long-term residence of the mobile population and thus increase the likelihood of having children. High housing prices mean that it is more difficult for the mobile population to integrate subjectively into the local area, and accordingly, their willingness to have children decreases.

On the other hand, housing prices are also a test of the economic capacity of mobile populations. The current direction of population mobility is mainly from rural to urban areas \citep{lewis1954economic}. Under the household registration system, which was born in the context of a dualistic economic system, this test is mainly in terms of job market disadvantages, financing constraints, and lack of public protection. First, mobile populations with the same endowments have less access to permanent jobs in the legitimate sector than local residents. The household registration system provides a natural barrier to entry into the job market, and self-employment becomes a rational choice for this group to cope with employment discrimination compared to wage employment\citep{banerjee1983experimental, roberts2001determinants, song2008social}. Second, the mobile population suffers from a very pronounced financial disincentive to start a business. Under the imperfect financial mechanism, their entrepreneurship tends to rely more on their own capital, which further reduces the possibility of capital circulation and expanded reproduction \citep{paulson2004entrepreneurship}. Third, the entrepreneurial risk and cost of living increase because the mobile population does not enjoy a series of social security brought by the household registration system. For example, \cite{oates1969effects} emphasizes the role of public services in a utility model of population migration, arguing that a good social security system will reduce expenditures in many ways and is an important driver of migration decisions. In summary, the series of thresholds imposed by the household registration system make the mobile population relatively less able to afford housing prices and lead to a greater reluctance to bear the various costs arising from childbearing.

There are still some gaps in academic research on fertility intentions of specific groups, and most of the benchmark indicators refer to absolute house prices. This paper takes the mobile population as the main research subject and relative house prices as the core explanatory variable, which provides a new research perspective and policy basis for pulling the benign population growth.

\section{Data sources and variable selection}
\begin{enumerate}[(1)]
\item Data sources.
\end{enumerate}

This paper selects data from the "Mobile Population Dynamic Monitoring Survey (CMDS)" organized by the National Health Care Commission in 2018 for empirical analysis. The survey covered 31 provinces (municipalities and autonomous regions) in mainland China, and the sample was selected from the migrant population who had stayed in the local area for more than one month. The average age of the sample is in the 15-59 years old range, and a stratified, multi-stage, large-scale PPS sampling method is adopted to investigate the development, individual characteristics, social integration, and employment of China's migrant population in detail, providing data support for social science research in the direction of labor economics and demography. The data related to urban control variables are obtained from the statistical yearbooks of various cities.

\begin{enumerate}[(2)]
\item Variable selection.
\end{enumerate}
1. Core variables

The explanatory variable of this paper is the fertility intention of the migrant population, and we choose the question "Do you have a fertility plan at the moment?" in the CMDS questionnaire. The core explanatory variable in this paper is the relative house price, which is the ratio of house price and household income in the city where the migrant population lives. Household income is measured by "what is your total monthly household income" in the CMDS questionnaire, which is a clearer indicator of the difficulty of acquiring property rights in the city than the relative house price constructed by per capita disposable income in the city.

2. Control variables screening and identification strategy

The control variables in this paper include both urban control variables and individual control variables. In existing studies on fertility, factors such as years of education, family size, age, and insurance participation are frequently found as demographic control variables. Variables such as urban health care level and disposable income per capita are frequently occurring urban control variables. In order to reduce the endogeneity problem arising from omitted variables in the research design phase, this paper selected 28 previously occurring variables from previous studies on fertility intentions and controlled for variables with strong explanatory power and wide coverage from the variables to be selected through a machine learning LASSO model to improve the estimation results\citep{dettling2014house, flavin2011owner, JingYA2016, LiYH2019}.
LASSO, as a form of penalized regression, is based on the principle of imposing certain constraints on the regression coefficients to avoid the introduction of too many explanatory variables in the equation, retaining those variables that have significant influence on the explanatory variables, which helps to control the screening of variables and obtain better prediction results. Its objective function is:
\newtheorem{theorem}{Theorem}
\begin{theorem}
\begin{eqnarray}\label{10}
\min _{\beta_{0}, \beta} \frac{i}{2 N}\left(\sum_{i=1}^{N}\left(y_{i}-\beta_{0}-x_{i}^{T} \beta\right)^{2}+\lambda \sum_{j=1}^{k}\left|\beta_{j}\right|\right)
\end{eqnarray}
\end{theorem}

In equation (1), $y_{i}$ is the dependent variable, $x_{i}$ denotes the independent variables including the core explanatory variables, beta denotes the regression coefficient, ${k}$ denotes the number of explanatory variables, lambda is a non-negative regularization parameter, and $\sum_{i=1}^{N}\left(y_{i}-\beta_{0}-x_{i}^{T} \beta\right)^{2}$ denotes the prediction error of the multiple regression results, which $\lambda \sum_{j=1}^{k}\left|\beta_{j}\right|$ corresponds to a penalty function excluding variables with weak explanatory power for the model in a given case so that the coefficients corresponding to these variables take the value of zero.

\begin{table}[htb]
\caption{Baseline regression results. }\label{tbl1}
\begin{tabular*}{\tblwidth}{@{} CCC@{} }
\toprule
Introduction to variables                     & Explanation of variables                                                                                                                   & Lasso factor \\
\midrule   
implicit variable                             & \begin{tabular}[c]{@{}c@{}}Intention to have children\\  (1 if you want to have children, 0 if you don't or don't want to)\end{tabular}    & -            \\
\midrule   
Core explanatory variables                    & \begin{tabular}[c]{@{}c@{}}Relative house prices \\ (house prices/household income)\end{tabular}                                           & -.0026868    \\
\midrule   
\multirow{6}{*}{Individual control variables} & Number of family members                                                                                                                   & -.1037128    \\
                                              & Number of years of education                                                                                                               & .0053022     \\
                                              & \begin{tabular}[c]{@{}c@{}}Health status \\ (consider yourself unhealthy for 1 and healthy for 4)\end{tabular}                             & .0049792     \\
                                              & \begin{tabular}[c]{@{}c@{}}Nature of household registration \\ (1 for resident, 0 for agricultural)\end{tabular}                           & .0028327     \\
                                              & \begin{tabular}[c]{@{}c@{}}Nature of work \\ (2 for civil servants, institutions, 1 for other institutions, 0 for unemployed)\end{tabular} & .0023393     \\
                                              & age                                                                                                                                        & -.0099228    \\
\midrule                                                 
\multirow{5}{*}{Urban control variables}      & Share of tertiary sector                                                                                                                   & .0003028     \\
                                              & Percentage of urban green space                                                                                                            & .0000128     \\
                                              & Book collection per capita                                                                                                                 & -.0023456    \\
                                              & Number of primary school teachers per capita                                                                                               & .0018002     \\
                                              & Number of physicians per capita                                                                                                            & -.0002457    \\
\bottomrule
\end{tabular*}
\end{table}
In this paper, 28 variables were selected for Lasso regression from both urban and individual perspectives, and 12 variables were selected from them for inclusion in the control variables after comprehensive consideration. Due to the problem of space, only the Lasso estimation results of the screening variables are reported in this paper.

\begin{enumerate}[(3)]
\item Descriptive statistics of variables.
\end{enumerate}
Table 2 reports the results of descriptive statistics for all variables in the study. It can be seen that about 12\% of the total sample of all women of reproductive age have the desire to have children, and that the average household income of two months of the mobile population can offset the price of a square meter of property, which has a large variance. The average number of family members in the sample is three, the average number of years of education is around 10 years, most of them are in agricultural households, and the average age is about 37 years.

\begin{table}[htb]
\caption{Baseline regression results. }\label{tbl2}
\begin{tabular*}{\tblwidth}{@{} CCCCCCC@{} }
\toprule
Introduction to variables                     & meaning                                      & Obs   & Mean     & Std. Dev. & Min      & Max      \\
\midrule 
implicit variable                             & Fertility intentions                         & 45593 & .1198664 & .3248082  & 0        & 1        \\
\midrule 
Core explanatory variables                    & Relative to house prices                     & 45593 & 2.115211 & 65.32412  & 0        & 24275.6  \\
\midrule 
\multirow{6}{*}{Individual control variables} & Number of family members                     & 45593 & 3.140605 & 1.210802  & 1        & 12       \\
                                              & Number of years of education                 & 45593 & 10.29905 & 3.451954  & 0        & 19       \\
                                              & health status                                & 45593 & 1.153993 & .4195443  & 1        & 4        \\
                                              & Nature of household                          & 45593 & .3121184 & .4633594  & 0        & 1        \\
                                              & Nature of work                               & 45593 & .8089934 & .5557937  & 0        & 2        \\
                                              & age                                          & 45593 & 37.0503  & 11.20359  & 15       & 90       \\
\midrule                                               
\multirow{5}{*}{Urban control variables}      & Share of tertiary sector                     & 45593 & 54.64655 & 11.21521  & 26.54    & 80.98    \\
                                              & Percentage of green space                    & 45593 & 41.83624 & 3.467361  & 23       & 93.81    \\
                                              & Book collection per capita                   & 45593 & 1.620107 & 1.532317  & .0625    & 8.741936 \\
                                              & Number of primary school teachers per capita & 45593 & 46.57175 & 19.34495  & 18.54151 & 163.3362 \\
                                              & Number of physicians per capita              & 45593 & 66.02081 & 22.85496  & 17.82246 & 130.3405 \\      
\bottomrule
\end{tabular*}
\end{table}

\section{Research design}
\begin{enumerate}[(1)]
\item Baseline model setting and research assumptions.
\end{enumerate}

1. Baseline model setting

In this paper, the explanatory variable fertility intention is dichotomous and models that can be used include ordinal Probit, binary Logit, etc. According to N. Nunn, L. Wantchekon (2011), the least squares estimation assumptions in the regression process are less and the results are more robust in statistical significance compared to methods such as likelihood estimation, and the OLS method is still recommended when the explanatory variable is a dummy variable or an ordered variable. In this paper, the Probit model is used as the main regression model, while the results of Linear Probability Model (LPM) are reported.
The basic expression for the Probit model is.

\begin{theorem}
\begin{eqnarray}\label{101}
\operatorname{Pr}\left(\mathrm{T}_{i j}=1 \mid h i, X\right)=F\left(h i, \beta_{1}\right)=\frac{\exp \left(\beta_{0}+\beta_{1} h i_{i j}+\beta X+\varepsilon_{i j, t}\right)}{1+\exp \left(\beta_{0}+\beta_{1} h i_{i j}+\beta X+\varepsilon_{i j, t}\right)}
\end{eqnarray}
\end{theorem}

In equation (2), ${T}_{i j}$ the intention of the mobile population to have children is a binary variable. If the mobile population ${i}$ chooses to give birth in the ${j}$ inflow city, the value is 1, otherwise it is 0. ${Pr}\left(\mathrm{T}_{i j}=1 \mid h i, X\right)$ is the probability that the mobile population ${i}$ has the intention to give birth in the inflow city ${j}$. ${F}$ is the cumulative distribution function of the standard normal. ${hi}$is house price to income ratio, X are other control variables, $\varepsilon_{i j, t}$ and are random disturbance terms.

The basic expression for the LPM model is:

\begin{theorem}
\begin{eqnarray}\label{102}
T_{i j}=\alpha+\beta h i_{i j}+X_{i j}^{\prime} \gamma+\varepsilon_{i j}, \varepsilon_{i j} \sim N\left(0, \quad \sigma^{2}\right)
\end{eqnarray}
\end{theorem}

In equation (3),$T_{i j}$ is the fertility intention of the mobile population;${hi}$ is the relative house price of mobile households, $X_{i j}^{\prime} \gamma$ is the individual, urban control variables. $\varepsilon_{i j}$ is the random disturbance term.

\begin{enumerate}[(2)]
\item Research hypothesis.
\end{enumerate}
Based on the benchmark regression, this paper proposes the following hypothesis on the transmission mechanism by which relative house price house prices affect the fertility intentions of the mobile population.
Hypothesis 1: There is a negative relationship between the relative house prices in the cities where the mobile population is located and their fertility intentions.

The level of housing prices is related to the willingness of the mobile population to integrate socially. As a concrete quantification of property rights, house prices are an important indicator of the many ways in which mobile people integrate into the society to which they move, and thus influence their willingness to stay. Since migration is a short-term decision, lowering the threshold of ownership is a key step in promoting long-term residence and thus increasing the likelihood of having children. High property prices mean that it is more difficult for mobile populations to integrate subjectively into the local area and, accordingly, their willingness to have children decreases.

Hypothesis 2: The sensitivity of fertility intentions of the mobile population to relative house prices is influenced by local infrastructure development (interaction effect).

That is, the preference of the mobile population for local property rights is partly due to local factors such as education and environment. When the conditions of infrastructure services are better in the inflowing place, the inflowing population is more sensitive to property prices, which is also reflected in their willingness to have children, and there is an interaction effect between the construction of infrastructure services on the willingness of the mobile population to have children and the sensitivity of property prices.
Hypothesis 3: The sensitivity of fertility intentions of the mobile population to house prices depends on the burden of property rights.

House prices are a test of the financial capacity of the mobile population. The fertility intentions of mobile populations under housing pressure are affected by house prices, while those without housing pressure are not necessarily sensitive to house prices. Housing expenses put enormous financial pressure on the mobile population, squeezing their ability to bear the costs of childbearing.

Of these, hypothesis one can be proved by the results of the benchmark regression.

In response to hypothesis two, this paper examines the possible moderating effects using an interaction term between infrastructure variables and relative house prices.The basic expression for the Probit model is.

\begin{theorem}
\begin{eqnarray}\label{10}
\operatorname{Pr}\left(\mathrm{T}_{i j}=1 \mid h i, X\right)=F\left(h i, \beta_{1}\right)=\frac{\exp \left(\beta_{0}+\beta_{1} h i_{i j}+\beta_{2} \inf _{i j}+\beta_{1} h i_{i j} \cdot \beta_{1} \inf _{i j}+\beta X+\varepsilon_{i, t}\right)}{1+\exp \left(\beta_{0}+\beta_{1} h i_{i j}+\beta_{2} \inf _{i j}+\beta_{1} h i_{i j} \cdot \beta_{1} \inf _{j j}+\beta X+\varepsilon_{i j, t}\right)}
\end{eqnarray}
\end{theorem}

In equation (4), ${T}_{i j}$ is the fertility intention of the mobile population, ${Pr}\left(\mathrm{T}_{i j}=1 \mid h i, X\right)$ is the probability of the mobile population ${i}$ having fertility intention in the inflow city ${j}$. ${F}$ is the cumulative distribution function of the standard normal. ${hi}$ is the house price to income ratio, and $\inf _{i j}$  is the infrastructure variable.${X}$ is other control variables, $\varepsilon_{i j, t}$ and are random disturbance terms.

In response to hypothesis three, this paper regresses the mobile population in groups according to ownership status. In the CMDS2018 questionnaire, there is no question related to the property ownership status of the mobile population, and this paper uses household housing expenditure as a proxy variable for property ownership status - the group with zero household housing expenditure is regressed in a separate group to examine the effect of property ownership on the mechanism of action.

\begin{enumerate}[(3)]
\item Endogeneity issues and robustness tests.
\end{enumerate}
The core explanatory variable in the model setting of this paper: fertility intention of the mobile population is influenced by a number of factors. In this paper, variables are screened by Lasso penalized regression in the variable selection stage, but it is still difficult to control all potential influencing factors. Meanwhile, high fertility intention of the mobile population in a region may raise the house price, creating a two-way causality problem and affecting the unbiased assumption of the estimation. Instrumental variables are effective methods in dealing with omitted variables and endogeneity problems due to two-way causality. Common instrumental variables in empirical studies targeting house prices include land development area, the product of long-term interest rate and land supply elasticity, and land sale price \citep{Peng2016, Zhangwei2018, chaney2012collateral, waxman2020tightening}. In this paper, we choose lagged land price as the instrumental variable of house price. Land sale prices directly affect house prices, which is consistent with the instrumental variable correlation hypothesis; meanwhile, the mobile population does not actively consider solving the land sale prices of previous years when considering fertility, which is consistent with the instrumental variable exogeneity hypothesis. To conform to the data structure of relative house prices, the lagged land price to lagged disposable income per capita ratio is constructed as the instrumental variable in this paper.

As above, this paper reports regression results for both IVProbit and IVLPM. In the two-stage least squares method, the hypothesis of weak instrumental variables can in principle be rejected by analyzing the first-stage F-statistic, and to ensure the robustness of the conclusions, this paper also chooses the limited information maximum likelihood (LIML) method, which is more accurate in predicting weak instrumental variables, for the regression. In addition, generalized distance estimation (GMM) is relatively more effective if the nuisance terms are heteroskedastic or autocorrelated. To prevent potential nuisance terms from affecting the unbiasedness of the estimates, generalized distance estimation was added for robustness testing with iterative generalized distance estimation (IGMM).

Although the sampling method for monitoring the mobility dynamics is scientific and the sample size is large, there may still be endogeneity problems arising from "self-selection" in the sample selection. A robustness test using propensity score matching can effectively alleviate this situation. In this paper, the 1:1 nearest neighbor matching method is used for analysis, and the average treatment effects of the 1:2 nearest neighbor matching method, radius matching and kernel matching are tested for robustness.

\section{Empirical process and discussion}
\begin{enumerate}[(1)]
\item Baseline regression results.
\end{enumerate}

\begin{table}[htb]
\caption{Baseline regression results. }\label{tbl1}
\begin{tabular*}{\tblwidth}{@{} CCCCCCCC@{} }
\toprule
                             & (1)         & (2)        & (3)         & (4)          & (5)         & (6)       & (7)        \\
                             & OLS\_1      & OLS\_2     & OLS\_3      & OLS\_4       & OLS\_5      & Probit\_1 & Probit\_2  \\
\midrule                              
houseincome                  & -0.00271*** & -0.00129*  & -0.00690*** & -0.00269***  & -0.00302**  & -0.0140** & -0.0218**  \\
                             & (0.000898)  & (0.000674) & (0.00219)   & (0.00102)    & (0.00118)   & (0.00573) & (0.0110)   \\
jobdata                      & -           & 0.00410    & -           & 0.00262      & 0.00211     & -         & -0.0192    \\
                             & -           & (0.00261)  & -           & (0.00272)    & (0.00278)   & -         & (0.0165)   \\
health                       & -           & 0.00642**  & -           & 0.00560*     & 0.00561*    & -         & -0.0249    \\
                             & -           & (0.00311)  & -           & (0.00321)    & (0.00327)   & -         & (0.0301)   \\
old                          & -           & -0.0100*** & -           & -0.00995***  & -0.00983*** & -         & -0.0687*** \\
                             & -           & (0.000193) & -           & (0.000200)   & (0.000202)  & -         & (0.00144)  \\
family                       & -           & -0.101***  & -           & -0.104***    & -0.107***   & -         & -0.696***  \\
                             & -           & (0.00178)  & -           & (0.00189)    & (0.00193)   & -         & (0.0182)   \\
edudata                      & -           & 0.00493*** & -           & 0.00531***   & 0.00555***  & -         & 0.0412***  \\
                             & -           & (0.000460) & -           & (0.000490)   & (0.000510)  & -         & (0.00344)  \\
hkdata                       & -           & 0.00408    & -           & 0.00305      & 0.000380    & -         & 0.0289     \\
                             & -           & (0.00310)  & -           & (0.00323)    & (0.00382)   & -         & (0.0229)   \\
\midrule                              
green                        & -           & -          & 0.000145    & 0.0000614    & 0.110***    & -         & 0.761***   \\
                             & -           & -          & (0.000451)  & (0.000422)   & (0.0324)    & -         & (0.263)    \\
pbook                        & -           & -          & -0.00114    & -0.00261*    & -0.0999***  & -         & -0.691***  \\
                             & -           & -          & (0.00152)   & (0.00135)    & (0.0288)    & -         & (0.231)    \\
pteacher                     & -           & -          & 0.000696*** & 0.00182***   & 0.0351***   & -         & 0.235***   \\
                             & -           & -          & (0.000179)  & (0.000165)   & (0.00958)   & -         & (0.0795)   \\
phospitalr                   & -           & -          & 0.0000226   & -0.000263*** & -0.00227*** & -         & -0.0146**  \\
                             & -           & -          & (0.0000951) & (0.0000872)  & (0.000714)  & -         & (0.00594)  \\
gdp                          & -           & -          & 0.00137***  & 0.000353*    & -0.00773*** & -         & -0.0536*** \\
                             & -           & -          & (0.000214)  & (0.000188)   & (0.00226)   & -         & (0.0176)   \\
\midrule                              
individual control variables & -           & $\surd$    & -           & $\surd$      & $\surd$     & -         & $\surd$    \\
Urban control variables      & -           & -          & $\surd$     & $\surd$      & $\surd$     & -         & $\surd$    \\
fixed effects                & -           & -          & -           & -            & $\surd$     & -         & $\surd$    \\
\midrule 
cons/cut1                    & 0.124***    & 0.759***   & 0.0201      & 0.688***     & -4.640***   & 1.152***  & 34.39***   \\
                             & (0.00214)   & (0.0137)   & (0.0208)    & (0.0236)     & (1.588)     & (0.0120)  & (13.02)    \\
\midrule                              
\textit{N}                   & 49453       & 49453      & 45593       & 45593        & 45593       & 49453     & 45593      \\
\textit{R2}                  & 0.000       & 0.157      & 0.003       & 0.162        & 0.170       & -         & -          \\
\bottomrule
\end{tabular*}
\end{table}

\begin{table}[htb]
\caption{Baseline regression results. }\label{tbl1}
\begin{tabular*}{\tblwidth}{@{} CCCCCC@{} }
\toprule
                             & (8)       & (9)      & (10)      & (11)      & (12)      \\
                             & TSLS      & IVProbit & LIML      & GMM       & IGMM      \\
\midrule                             
Relative to house prices     & -0.0442** & -0.222** & -0.0442** & -0.0442** & -0.0442** \\
\multicolumn{1}{l}{}         & (0.0200)  & (0.0965) & (0.0200)  & (0.0200)  & (0.0200)  \\
\midrule
Individual control variables & $\surd$   & $\surd$  & $\surd$   & $\surd$   & $\surd$   \\
Urban control variables      & $\surd$   & $\surd$  & $\surd$   & $\surd$   & $\surd$   \\
\midrule
N                            & 45593     & 45593    & 45593     & 45593     & 45593     \\
R2                           & 0.121     & -        & 0.121     & 0.121     & 0.121     \\
\midrule
One-stage regression results & \multicolumn{5}{c}{F=47.9163, p=0.0000}                  \\
\bottomrule
\end{tabular*}
\end{table}

Table 3 shows the results of the baseline regression model. Models (1) and (2) are the results of the LPM regressions. Model (1) indicates that the regression coefficient of the house price to income ratio on fertility intention is -0.00271, which is significant at 1\% level, conditional on the inclusion of other control variables, indicating that there is a significant negative effect of higher house prices on the fertility intention of the mobile population, and the higher the ratio of house prices to the household income of the mobile population, the lower their fertility intention. Model (2) includes both urban and individual control variables, and the regression coefficient does not change significantly (-0.00269) compared to the regression coefficient under the condition of no control variables included.

The estimated results of the Probit model LPM are similar, i.e., there is a significant negative effect of high house prices on the fertility intentions of the mobile population. Model (3) indicates that the regression coefficient of house price to income ratio on fertility intention is -0.0140 when no other control variables are included. model (4) The regression coefficient of relative house price on fertility intention of the mobile population is relatively lower (-0.00987) and slightly less significant (5\%) when individual control variables are included. Model (5) has a relatively higher effect of fertility intentions of the mobile population (-0.0344) when urban control variables are included. Model (6), which includes both urban and individual control variables, has a slightly higher regression coefficient (-0.0195) compared to the regression coefficient under the condition that no other control variables are included.

Table 4 depicts the predicted results after adding IV to the model, and the results of 2SLS in model (7) show some increase in the absolute value of the coefficients compared to the results of the LPM regression, indicating that the underlying endogeneity problem tends to underestimate the effect of house prices on the fertility intentions of the mobile population. For each female age-eligible mobile individual, each 100 percent increase in the relative house price to income ratio will reduce the intention to have children locally by 4.42 percent.

According to the predictions of model (8) IVProbit, the regression coefficient of relative house prices is -0.222, which is significant at 5\% level, indicating that given constant individual and urban characteristics, the marginal effect of relative house prices on labor force fertility intentions compared to female age mobile individuals without fertility intentions is exp(-0.222) , i.e. for every 100\% increase in relative house prices, each female age mobile individual will have a 22.2 percent reduction in fertility intentions.

According to model (9), the regression results for the two-stage least squares regression and the limited information maximum likelihood estimation are highly similar, again demonstrating that there is no weak instrumental variable problem in the empirical evidence. The results of models (10) and (11) using generalized distance estimation and iterative generalized moments estimation are similar to the 2SLS results as well, indicating that the heteroskedasticity problem of potential nuisance terms does not significantly interfere with the empirical study and that the estimation results are robustly keyed.

In summary, hypothesis 1: "The relative house prices in the cities to which the mobile population flows are negatively related to their fertility intentions" is confirmed.

\begin{enumerate}[(2)]
\item propensity score matching
\end{enumerate}
Propensity score matching can effectively mitigate the endogeneity problem caused by sample selection bias. In this paper, we use the matched sample for analysis to estimate the net effect of matched relative house prices on the fertility intentions of the mobile population after passing a common support hypothesis test and a stationarity test. Table 5 shows the PSM estimation results, and the average treatment effect of all four matching methods is around -0.24, controlling for individual characteristics of the sample, which is the same conclusion as the baseline regression results: i.e., there is a significant negative relationship between relative house prices and fertility intentions of the mobile population.

\begin{table}[htb]
\caption{Baseline regression results. }\label{tbl1}
\begin{tabular*}{\tblwidth}{@{} CCCCCCCCC@{} }
\toprule
\multirow{2}{*}{Matching Status} & \multicolumn{2}{c}{1:1 Matching} & \multicolumn{2}{c}{1:2 match} & \multicolumn{2}{c}{radius matching} & \multicolumn{2}{c}{nuclear matching} \\
                                 & ATT               & t            & ATT             & t           & ATT                & t              & ATT                 & t              \\
\midrule                                 
pre-match                        & -.0159***         & -5.26        & -.0159***       & -5.26       & -.0159***          & -5.26          & -.0159***           & -5.26          \\
post-match                       & -.0241***         & -4.08        & -.0234***       & -4.36       & -0.217***          & -4.25          & -0.224***           & -4.35          \\
\bottomrule
\end{tabular*}
\end{table}

\begin{enumerate}[(3)]
\item Moderating effects of infrastructure
\end{enumerate}
The effect of relative house prices on the fertility intentions of the mobile population was estimated above, but is the effect of house price to income ratio on the measures always direct? Table 6 reports the relationship between the moderating effects of education and greening-related facilities on house prices and fertility intentions in mobile inflow locations.

Models (12) and (14) put in only the interaction term of relative house prices and infrastructure public services, and (13) and (15) include urban, individual control variables. In particular, models (12) and (13) are the interactions of book collection per capita and relative house price, with regression coefficients of -0.00439 and -0.00851 without and with the inclusion of control variables, respectively, which are significant at 1\% and 5\% water level. Models (14) and (15) are the interactions between urban green area and relative house prices, with coefficients of -0.000263 and 0.00329 without and with control variables, respectively, significant at 5\% and 10\% levels. All these results suggest that there is a moderating effect of the level of infrastructure services on relative house prices. The preference of the mobile population for local property rights is to some extent due to local education and environment, and the sensitivity of the mobile population's fertility intention to relative house prices is affected by local infrastructure development, and hypothesis two is confirmed.

\begin{table}[htb]
\caption{Baseline regression results. }\label{tbl1}
\begin{tabular*}{\tblwidth}{@{} CCCCC@{} }
\toprule
                             & (13)        & (14)      & (15)        & (16)      \\
                             & Probit\_1   & Probit\_2 & Probit\_3   & Probit\_4 \\
\midrule                             
hosbook                      & -0.00439*** & -0.00851* & -           & -         \\
                             & (0.00130)   & (0.00477) & -           & -         \\
hosgreen                     & -           & -         & -0.000263** & 0.00329*  \\
                             & -           & -         & (0.000116)  & (0.00172) \\
\midrule                             
Individual control variables & ×           & $\surd$   & ×           & $\surd$   \\
Urban control variables      & ×           & $\surd$   & ×           & $\surd$   \\
\midrule
N                            & 45593       & 45593     & 45593       & 45593     \\
\bottomrule
\end{tabular*}
\end{table}

\begin{enumerate}[(3)]
\item The effect of housing stress on the mechanism of action
\end{enumerate}

\begin{table}[htb]
\caption{Baseline regression results. }\label{tbl1}
\begin{tabular*}{\tblwidth}{@{} CCCCCCC@{} }
\toprule
                             & (17)       & (18)       & (19)      & (20)      & (21)     & (22)      \\
                             & OLS\_1     & OLS\_2     & OLS\_3    & OLS\_4    & IV\_1    & IV\_2    \\
\midrule                             
houseincome                  & -0.00231** & -0.00287** & -0.00189* & -0.00308* & -0.0292* & -0.0355  \\
                             & (0.00105)  & (0.00122)  & (0.00106) & (0.00157) & (0.0172) & (0.0310) \\
\midrule                             
individual control variables & -          & -          & $\surd$   & $\surd$   & $\surd$  & $\surd$  \\
Urban control variables      & -          & -          & $\surd$   & $\surd$   & $\surd$  & $\surd$  \\
\midrule
\textit{N}                   & 11889      & 37564      & 11038     & 34555     & 10973    & 34417    \\
\textit{R2}                  & 0.000      & 0.000      & 0.160     & 0.163     & 0.134    & 0.143    \\
\bottomrule
\end{tabular*}
\end{table}

In the hypothesis of this paper, the difficulty of acquiring relative house price as a measure of housing ownership is negatively related to the fertility intention of the mobile population. Another intuitive manifestation of this logic is that the transmission mechanism becomes more pronounced when the household housing expenditure of the mobile population is higher. In the CMDS questionnaire, household housing expenditure mainly includes rent and mortgage. In this paper, the mobile population is grouped into regressions according to whether the housing expenditure is zero or not. Table 7 reports the regression results of multiple methods for both groups. It is clear that the group with high household housing expenditure among the mobile population is more sensitive to housing prices, and hypothesis three is confirmed.

\begin{enumerate}[(5)]
\item Heterogeneity analysis
\end{enumerate}
The overall estimation of the effect of relative house prices on the fertility intention of the mobile population is presented above, but the conclusions drawn in the context of a large sample size and wide coverage are slightly generalized, and there is a certain need for heterogeneity analysis based on some characteristics of the sample. Among the methods for comparing the differences in coefficients between groups in group regressions, common tests include the introduction of cross terms, the seemingly uncorrelated model test, and the Fisher's combination test. The seemingly uncorrelated model assumption is relatively loose, allowing two groups of disturbance terms to be correlated with each other, which is more consistent with the research hypothesis of this paper.

The results are reported with estimates for subgroups below the median on the left and for subgroups above the median on the right. In Table 8, models (20) and (21) report regression results based on grouping by sample household size (3 persons), models (22) and (23) report regression results based on grouping by whether or not they have higher education, and models (24) and (25) report regression results based on grouping by age (30 years). According to the results, the fertility intentions of smaller family size, higher education level, and lower age of female mobile population of school age are more sensitive to house prices. The reasons for larger family size may include supporting the elderly or already having children, in either case, there is a certain burden on the family economy, and also the fertility intention of the second child is significantly lower compared to the first child (Mao et al.,2013). The mobile group with high education level tends to consider settling in the inflow area, and their fertility intention will be more sensitive to the house price.

\begin{table}[htb]
\caption{Baseline regression results. }\label{tbl1}
\begin{tabular*}{\tblwidth}{@{} CCCCCCC@{} }
\toprule
                           & (23)            & (24)          & (25)                    & (26)                   & (27)        & (28)        \\
                           & \multicolumn{2}{c}{Family size} & \multicolumn{2}{c}{Number of years of education} & \multicolumn{2}{c}{age}   \\
\midrule                           
hosin                      & -0.00507***     & -0.000562     & -0.00143**              & -0.0189***             & -0.0156***  & -0.00127*   \\
                           & (0.00168)       & (0.000498)    & (0.000704)              & (0.00417)              & (0.00336)   & (0.000680)  \\
\midrule                           
N                          & 24674           & 20919         & 36338                   & 9255                   & 13255       & 32338       \\
R2                         & 0.221           & 0.038         & 0.132                   & 0.183                  & 0.136       & 0.115       \\
\midrule
Differences between groups & \multicolumn{2}{c}{0.005}       & \multicolumn{2}{c}{0.019}                        & \multicolumn{2}{c}{0.015} \\
\midrule
P-value                    & \multicolumn{2}{c}{0.000}       & \multicolumn{2}{c}{0.000}                        & \multicolumn{2}{c}{0.000} \\
\bottomrule
\end{tabular*}
\end{table}

\section{Conclusion}
The size of the country's migrant population continues to expand, and the issue of social integration is of wide concern. At the same time, the fertility rate is significantly lower than the level of generational change required to achieve population stability and is close to the internationally recognized "low fertility trap" threshold of 1.3. Promoting the willingness of the mobile population to have children and creating a stable source of manpower for economic development is one of the ways to optimize the dual economic structure between urban and rural areas and promote China's advancement to a high-income country. The exploration of the fertility intention of the mobile population from the perspective of housing prices is not only related to the coordinated development of the regional economy, the reduction of the gap between the rich and the poor and the alleviation of social conflicts, but also can optimize the allocation of regional labor endowment structure and promote the stable and healthy development of China's economy. This paper tries to provide a new practical perspective for the mainline institutional change and balanced economic development in China's economic transition stage.

This paper uses OLS and Probit models to estimate the effect of house prices on the fertility intentions of the mobile population based on data from the 2018 National Mobility Dynamics Monitoring Survey. The lagged land sale price is used as an instrumental variable of house price to mitigate the potential endogeneity problem. The results show that for every 100\% increase in the ratio of house price to household income of the migrant population, the fertility intention of the female migrant population of the right age in the inflow area will decrease by 4.42\%, and the fertility intention of each female migrant individual of the right age will decrease by 22.2\%, i.e. the marginal effect of relative house price on labor force fertility intention is exp(-0.222); the sensitivity of the fertility intention of the migrant population to house price is affected by the moderating effect of infrastructure construction in the inflow area. The sensitivity of fertility intentions of the mobile population to house prices is influenced by the moderating effect of infrastructure construction in the inflow area. Female migrants with lower age, smaller family size and higher education have higher fertility intention in the inflow area. In summary, the paper discusses the following components.

\bibliography{cas-refs}

\end{document}